\documentclass[useAMS,usenatbib,usegraphicx,onecolumn]{mn2e}

\title{Polarization of the Broad H$\alpha$ Wing in Symbiotic Stars} 
 
\author[Yoo, Bak \& Lee]{Jerry Jaiyul Yoo$^1$ 
\thanks{E-mail: jaiyul@astronomy.ohio-state.edu} 
\thanks{Present address : Department of Astronomy, The Ohio State University, 
140 W. 18th Ave. Columbus, OH 43210} 
, Jih-Yong Bak$^2$ and 
Hee-Won Lee$^3$\\ 
$^1$ Astronomy Program, School of Earth and Environmental Sciences,  
Seoul National University, Seoul 151-742, Korea. \\  
$^2$ Department of Astronomy, Yonsei University, 134 Shinchon, 
Seodaemunku, Seoul 120-749, Korea\\ 
$^3$ Department of Geoinformation Sciences, Sejong University,  
Seoul 143-747, Korea} 
 
\date{Accepted 2002 June 6. Received 2002 June 6; in original form 
2002 April 2} 
 
\begin{document} 
\maketitle 
 
\begin{abstract} 
In many symbiotic stars there appear broad wings around  
H$\alpha$, of which the formation mechanisms proposed thus far include
a fast outflow, motion of the inner accretion disc, electron scattering 
and Raman scattering of Ly$\beta$.  
We adopt a Monte Carlo technique to simulate the Raman scattering of 
ultraviolet photons 
that are converted into optical photons around H$\alpha$, forming 
broad wings, and compute its polarization. 
Noting that many symbiotic stars exhibit a bipolar nebular morphology 
and polarization flip in the red-wing part of the Raman scattered  
O~{\tiny{VI}}  
features, we assume that the neutral scattering region is composed of the 
two components. The first component is a static cylindrical shell with 
finite thickness; and the second component is a finite slab that is moving 
away with velocity $v_p=100{\rm\ km\ s^{-1}}$ along the symmetry axis of the 
first component. The cylindrical shell component yields polarization 
in the direction parallel to the cylinder axis. The strongest polarization 
is obtained in the limit where the height of the cylinder approaches zero  
and the scattering region effectively becomes a circular ring. As the height 
of the cylinder increases, the resultant polarization decreases and 
becomes negligible in the limit of the infinite cylinder. The polarization 
near the line-centre is weaker than in the far wing regions because of the 
large Rayleigh scattering numbers due to the large scattering cross sections 
near the line centre. The receding polar scattering component produces 
strong polarization in the direction perpendicular to the cylinder axis.  
In the presence of a Ly$\beta$ emission-line component 
with an equivalent width 
$\sim 0.5{\rm\ \AA}$, the polarized flux exhibits a local maxima at 
$\lambda=6578{\rm\ \AA}$ that corresponds to the receding velocity with 
$6.4 v_p$ relative to H$\alpha$. When the both scattering components co-exist, 
the polarization is characterized by weak parallel polarization near 
the line-centre and strong perpendicular polarization in the red part. 
We discuss the observational implications of our computation. 
\end{abstract} 
 
 
\section{Introduction} 
Symbiotic stars are generally known to be interacting binaries 
of a mass losing giant and a hot white dwarf surrounded by an ionized 
nebula which is responsible for various prominent emission lines \citep{ken}. 
In the spectra of symbiotic stars, there exist very unique 
emission features around $6830 {\rm\ \AA}$ and $7088{\rm\ \AA}$. 
\citet{sch1} proposed that they are formed via Raman scattering 
by atomic hydrogen of O~{\tiny{VI}} 1032, 1038 doublet.  
When an O~{\tiny{VI}} line photon  
that is more energetic than Ly$\alpha$  is incident upon a hydrogen atom  
in its ground $1s$ state, it subsequently 
re-emits an outgoing photon with de-excitation into either $1s$ state 
or $2s$ state. In the first case, we have a Rayleigh scattering 
process with no 
change in the frequency in the rest frame of the scatterer. The second case 
corresponds to a Raman scattering process, from which we obtain 6830  
and 7088 line photons from incident O~{\tiny{VI}}~1032 and 1038 
photons, respectively. 
 
The scattering cross-sections associated with these Raman processes 
are of order $\sigma\sim 10^{-23}{\rm\ cm^2}$  
\citep[e.g.][]{sch1,nuss1,sad,lee1}, and therefore the operation 
of the O~{\tiny{VI}} Raman scattering requires the existence of both  
a highly ionized 
O~{\tiny{VI}} emission region and a fairly extensive neutral 
scattering region with H~{\tiny{I}} column density  
$N_{HI}\sim10^{23}\rm\ cm^{-2}$. 
The scattering cross section is very sensitive to the wavelength  
of the incident radiation in such a way that it  
steeply increases as the wavelength approaches 
the resonance wavelength of the scatterer.  
Therefore, continuum photons with wavelength 
near that of the Ly$\beta$ will be scattered in a neutral 
hydrogen region with much lower H~{\tiny{I}} column density than that is 
required for O~{\tiny{VI}} Raman scattering. 
 
\citet{vanw} showed that most symbiotic stars  
in the southern hemishpere exhibit very broad wings around H$\alpha$.  
A similar 
result was reported for those in the northern hemisphere by \citet{ivi}. 
The full width at zero intensity often exceeds 
$2000{\rm\ km\ s^{-1}}$. Similar broad wings also appear in young 
planetary nebulae, including M2-9, Mz3, and IC~4997 
\citep[e.g.][]{mir,arr}.  
In particular, for M2-9, \citet{bal} reported the existence of very 
broad H$\alpha$ wings with a width $\sim 10^4{\rm\ km\ s^{-1}}$. 
Remarkably, \citet{sel} presented 
the {\it Very Large Telescope} (VLT)
spectrum of the symbiotic star RR~Tel which shows H$\alpha$  
wings extending up to $12000{\rm\ km\ s^{-1}}$. 
According to \citet{vand}, post-asymptotic giant branch (post-AGB)
stars also exhibit broad H$\alpha$ wings. 
 
There have been many theoretical suggestions for formation mechanism
of the broad H$\alpha$ wings.
These include fast outflows \citep[e.g.][]{scm}, 
electron scattering \citep[e.g.][]{ber}, the accretion disc 
\citep[e.g.][]{qui} and the Raman scattering of Ly$\beta$. 
In this paper, we will not consider all those theoretical models but will  
focus only on the H$\alpha$ wing formation via the Raman scattering  
of Ly$\beta$. 
 
Nussbaumer et al. (1989) discussed the astrophysical importance 
of Raman scattering by atomic hydrogen and proposed that broad H$\alpha$  
wings can be formed through the Raman scattering of Ly$\beta$. This 
idea has been applied to young and compact planetary nebula IC~4997 
by \citet{lee3}. They computed the wing profile formed through 
Raman scattering of flat continuum around Ly$\beta$ and obtained 
an excellent agreement with the observed profile. 
Similar success has been achieved for many symbiotic stars by \citet{lee2}. 
 
The existence of a thick H~{\tiny{I}} region in symbiotic stars and young  
planetary nebulae can also be inferred from the  
He~{\tiny{II}} Raman scattered features. Because He~{\tiny{II}} is 
also a single electron atom, the 
wavelengths associated with the transitions from $n=2m$ state to the 
$2s$ state are very close to those corresponding to the $m$th Lyman series 
of hydrogen. Therefore, He~{\tiny{II}} Raman scattering requires much less  
H~{\tiny{I}} column density $N_{HI}\sim 10^{20}{\rm\ cm^{-2}}$  
than O~{\tiny{VI}} 1032, 1038 doublet.  
The He~{\tiny{II}} Raman-scattered 
lines have been found in the spectra of RR~Tel 
and He 2-106 \citep[e.g.][]{van,lee4} and the young planetary nebulae  
M2-9 (Lee et al. 2001), which highlights the plausibility of the 
operation of the Ly$\beta$ Raman scattering in these objects. 
Therefore, in this work, we will consider Raman scattering of 
Ly$\beta$ in neutral regions with $N_{HI}=10^{20}-10^{23}\rm\ cm^{-2}$. 
 
Spectropolarimetry will provide valuable information on the H$\alpha$ 
wing formation mechanism.  Raman-scattered radiation consists 
of purely scattered radiation without any mixing of direct radiation, 
and therefore Raman-scattered radiation can be strongly polarized. 
Polarization is also dependent on the scattering geometry. 
Spectropolarimetry may reveal the circumstellar matter distribution 
and be useful to understand the mass-loss process of the giant component. 
 
In this paper, we will compute the profile and the polarization of  
H$\alpha$ wings that are assumed to be formed through Raman scattering 
of Ly$\beta$. In section 2, we will briefly introduce the basic atomic 
physics relevant to the H$\alpha$ wing formation and describe the 
Monte Carlo procedure. In section 3, we present our main results. In the 
final section we discuss our results and observational implications.

\section{H$\alpha$ Wing Formation through Raman Scattering} 
 
\subsection{Branching Ratio} 
\begin{figure} 
\centerline{\includegraphics[width=8cm]{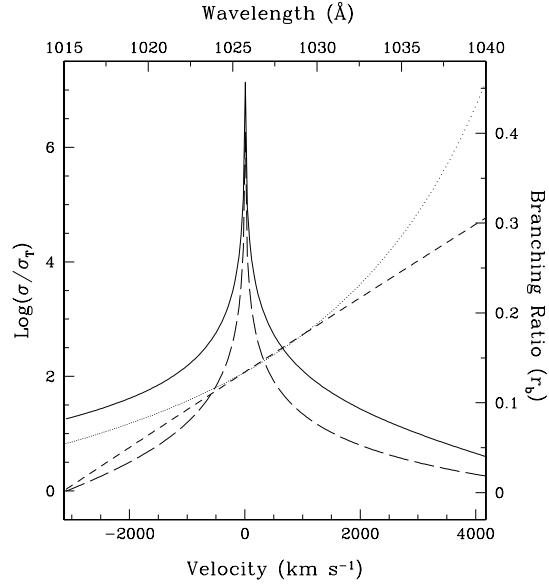}} 
\caption{Branching ratio of the Raman scattering to the Rayleigh scattering. 
The solid and the long dashed lines represent the differential cross sections 
of Rayleigh and Raman scatterings in units of Thomson scattering cross 
section, respectively. The branching ratio of the Raman scattering to the 
Rayleigh scattering and its first-order approximation are shown in the dotted  
and the short dashed lines, respectively.} 
\label{branch} 
\end{figure} 
A Ly$\beta$ photon incident on a scattering region consisting of atomic 
hydrogen undergoes either Rayleigh scattering or Raman scattering. In  
particular, in the latter case the scattering H atom de-excites to 
the $2s$ state resulting in an outgoing H$\alpha$ photon.  
Owing to the scattering incoherency, if an incident photon 
near the Ly$\beta$ line centre is Raman-scattered, then the outgoing 
photon finds itself in the H$\alpha$ wing region much further from the 
line centre, by a factor of 6.4, than from the line centre of 
Ly$\beta$ (e.g. Schmid 1989, Nussbaumer et al. 1989). Lee \& Hyung (2000) 
emphasized 
that the profile of H$\alpha$ wings formed by Raman scattering of 
Ly$\beta$ 
will be proportional to $\Delta \lambda^{-2}$, when the Raman 
scattering optical depth is small. However, the detailed conversion 
process from a ultraviolet photon into an optical photon is dependent on the 
branching ratio of the Rayleigh scattering and Raman scattering cross 
sections (e.g. Schmid 1996). In this subsection, we illustrate a basic 
calculation that provides an approximate branching ratio near the 
Ly$\beta$ 
line centre. 
 
The branching ratio of 
the Rayleigh process to the Raman process is well known in the literature,  
which is around 1:7. However, this ratio is dependent on the wavelength of  
the incident photon.  
 
The differential cross-section for Rayleigh scattering is given by 
\begin{equation} 
{\left( d\sigma \over d\Omega \right)}_{\rm Ray} = r_0^2  
\left| {1\over m\hbar}\sum_I  
\left({\omega({\bf p}\cdot{\mbox{\boldmath$\epsilon$}}^{\alpha'})_{AI} 
({\bf p}\cdot{\mbox{\boldmath$\epsilon$}}^\alpha)_{IA} 
\over \omega_{IA}(\omega_{IA}-\omega)}- 
{\omega({\bf p}\cdot{\mbox{\boldmath$\epsilon$}}^\alpha)_{AI} 
({\bf p}\cdot{\mbox{\boldmath$\epsilon$}}^{\alpha'})_{IA} 
\over \omega_{IA}(\omega_{IA}+\omega)}\right)\right|^2, 
\label{rayleigh} 
\end{equation} 
and the Raman scattering differential cross section is 
\begin{equation} 
{\left( d\sigma \over d\Omega \right)}_{\rm Ram} =  
r_0^2\left({\omega'\over\omega}\right) 
\left| {1\over m\hbar}\sum_I  
\left({({\bf p}\cdot{\mbox{\boldmath$\epsilon$}}^{\alpha'})_{BI} 
({\bf p}\cdot{\mbox{\boldmath$\epsilon$}}^\alpha)_{IA} 
\over \omega_{IA} -\omega}+ 
{({\bf p}\cdot{\mbox{\boldmath$\epsilon$}}^\alpha)_{BI} 
({\bf p}\cdot{\mbox{\boldmath$\epsilon$}}^{\alpha'})_{IA} 
\over \omega_{IA}+\omega'}\right)\right|^2, 
\label{raman} 
\end{equation} 
where the subscript $A$ stands for the initial quantum state of the 
scatterer, $I$ for the intermediate state and $B$ for the final state. 
We have  $A=1s, B=2s$ for the Raman process and $A=B=1s$  
for the Rayleigh process \citep{sak}.  
Here, $\omega_{IA}=\hbar^{-1}(E_I-E_A)$ is the angular frequency for the 
transition between the initial state $A$ and the intermediate state $I$, 
$r_0$ is the classical electron radius, and 
${\textit{\boldmath$\epsilon$}}^\alpha$ represents 
the polarization vector of the incident photon. 
The primed quantities correspond to those associated with 
the outgoing photon.  
In particular, the energy conservation relation for the Raman scattering 
case gives 
\begin{equation} 
\omega' = \omega - \omega_\alpha, 
\end{equation} 
where $\omega_\alpha$ represents the angular frequency of the 
Ly$\alpha$ transition. 
We are interested only in scatterings occurring in the 
wing regime far from the resonance and therefore omit the damping term  
in the denominator of each transition matrix element.  
 
For an incident photon around Ly$\beta$, $\omega \simeq \omega_\beta$ 
and the cross sections for both scattering processes are predominantly  
contributed by the first term in the summation corresponding to 
$I=3p$ in equations~(\ref{rayleigh}) and ~(\ref{raman}). 
Since the angular distribution 
of the scattered radiation is the same for both processes, we can write  
the branching ratio $r_b$ of the cross-sections for Rayleigh scattering 
and Raman scattering as
\begin{equation} 
r_b(\omega)=\left({\omega'\over\omega^3}\right)  
\left|{A_1/\Delta\omega-A_2 \over A_3/\Delta\omega-A_4} 
\right|^2  
\simeq\left( {\omega_\beta-\omega_\alpha \over \omega_\beta} 
\right)\left( {A_1\over {\omega_\beta A_3}}\right)^2   
\left[ 1+\left( {\omega_\beta\over \omega_\beta-\omega_\alpha} 
-3+{{2\omega_\beta A_2}\over A_1}-{{2\omega_\beta A_4}\over A_3} 
\right) {\Delta\omega\over \omega_\beta} \right], 
\label{ratio} 
\end{equation} 
where we set $\omega=\omega_\beta + \Delta\omega$,  
$|\Delta\omega|\ll\omega_\beta$. 
 
The matrix elements that appear in equation~(\ref{ratio}) are explicitly given by 
\begin{eqnarray} 
A_1 &=& <2s\parallel p\parallel 3p><3p\parallel p\parallel 1s>, \nonumber \\ 
A_2 &=& \sum_{I\ne 3p} {<2s\parallel p\parallel I><I\parallel p\parallel 1s> 
\over \omega_{I1} -\omega_\beta-\Delta\omega}+ 
\sum_{I\ne 3p}{<2s\parallel p\parallel I><I\parallel p\parallel 1s> 
\over \omega_{I1}+(\omega_\beta-\omega_\alpha+\Delta\omega)}, \nonumber \\ 
A_3 &=& {{|<3p\parallel p\parallel 1s>|^2} \over {\omega_\beta}} , \nonumber \\ 
A_4 &=& \sum_{I\ne 3p} {|<I\parallel p\parallel 1s>|^2 
\over \omega_{I1}(\omega_{I1} -\omega_\beta-\Delta\omega)}- 
\sum_{I\ne 3p}{|<I\parallel p\parallel 1s>|^2 
\over \omega_{I1}(\omega_{I1}+\omega_\beta+\Delta\omega)}.  
\end{eqnarray} 
It is noted that the summation extends to the continuum states, where 
the sum over all continuum states should be written in an integral form. 
 
The explicit representations of the matrix elements are found in the literature 
\citep[e.g.][]{kar,sas,rel} 
and a straightforward numerical computation gives 
$A_1/(\omega_\beta A_3)=0.9268$, 
$\omega_\beta A_4/A_3= 17.91$, 
$\omega_\beta A_2/A_1= -30.37$. 
These results are directly substituted into equation~(\ref{ratio}) to yield 
\begin{equation} 
r_b(\omega)\simeq 0.1342 -12.50\Delta\omega/\omega_\beta. 
\label{app} 
\end{equation} 
 
In Fig.~\ref{branch} we present the branching ratio of the Raman scattering  
cross-section to the Rayleigh cross section.  
The short-dashed line represents the approximate 
result given in equation~(\ref{app}), whereas the dotted line stands for the result 
obtained by using the whole Kramers-Heisenberg relation. 
The approximate relation gives an error of less than 10 per cent in the range 
$\Delta V \le 1000 {\rm\ km\ s^{-1}}$. 
This implies that the contribution from other $np$ states is not negligible. 
From Fig.~\ref{branch} we see that the branching ratio varies in a rather large 
range from 0.05 to 0.45 in the wavelength interval from 1015{\rm\ \AA} to 
1040 \AA. Since the broad feature around 7088{\rm\ \AA} which is Raman-scattered 
O~{\tiny{VI}} 1038 is often found in the spectra of symbiotic stars, it is 
necessary to investigate the Raman conversion efficiency for a broad range 
of the branching ratios. 
 
\subsection{Scattering Geometry} 
\begin{figure} 
\centerline{\includegraphics[width=8cm]{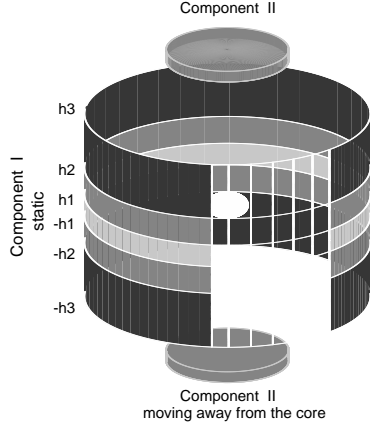}} 
\caption{Schematic view of the scattering geometry. The central core 
represents the emission region. The static cylindrical shell denoted 
by Component~I is composed of three layers stratified from the 
equatorial region. The layers are characterized by different  
H~{\tiny{I}} densities and heights $h_1,h_2,h_3$. The circular slabs 
(Component~II) with homogeneous H~{\tiny{I}} density are located in 
the polar direction, moving away from the core.} 
\label{geometry} 
\end{figure} 
Many symbiotic stars exhibit a bipolar nebular morphology, 
which is characterized by the accumulation of circumstellar materials   
near the equatorial plane. Many studies of Raman scattered O~{\tiny{VI}} lines 
have demonstrated that the neutral scattering region can be associated with 
the cool component, while the ionized region must be close to the hot 
component \citep[]{har2,sch5,sch4}.  
 
However, the Raman scattering of 
Ly$\beta$ becomes significant in neutral regions with  
$N_{HI}\ge 10^{20}\rm cm^{-2}$, which is two orders of magnitude 
smaller than the column density required for the operation of 
O~{\tiny{VI}} Raman scattering. 
This implies that the H$\alpha$ wing formation through Ly$\beta$  
Raman scattering should occur in much more extended regions than 
those that can be probed by Raman-scattered O~{\tiny{VI}} features. 
This is consistent with the fact that most symbiotic stars exhibit 
broad H$\alpha$ wings whereas only about half of them show 
Raman-scattered O~{\tiny{VI}} features 
(van Winckel et al. 1993, Ivison et al. 1994).
Furthermore, broad H$\alpha$ wings are also found in young 
planetary nebulae and post-AGB stars, which appear to have common 
characteristics including bipolar morphology and composite 
emission-line profiles \citep{arr}. It is still uncertain that all these 
objects are binary systems, whereas symbiotic stars are generally  
regarded as binary systems. In view of these circumstances, we adopt 
a simple neutral scattering region that is approximated by a 
cylindrical shell. 
 
Fig.~\ref{geometry}  
shows a schematic view of the 
scattering geometry.  
We just assume that the neutral hydrogen component forms a finite 
cylindrical shell (Component~I in Fig.~\ref{geometry}) 
with the centre coinciding with the Ly$\beta$ emission 
source. The cylindrical shell is assumed to be of constant H~{\tiny{I}} density 
and hence characterized by radius $r$ and height $h$. 
 
In the spectropolarimetric data presented by \citet{har}, 
the Raman-scattered O~{\tiny{VI}} 6830 and 7088 features exhibit polarization 
flip in the red-wing part. \citet{sch3} interpreted the polarization flip 
in the red-wing part by scattering in the slow stellar wind from the 
giant component. Another interpretation has been proposed by \citet{lee5}, 
who assumed the existence of an accretion disc around 
the white dwarf component via gravitational capture of the slow stellar 
wind around the giant component. According to their model, the emission 
region is formed around the accretion disc, and the main double peaks 
shown in the Raman-scattered features are attributed to the disc 
motion viewed in the equatorial direction, where the main scattering 
region is formed around the giant component. In order to explain 
the polarization flip in the red-wing part, there 
has to be another scattering component that is moving away from the 
emission region in the direction perpendicular to the equatorial plane. 
 
In this work, we will also consider that H$\alpha$ wings are significantly 
contributed by this polar scattering region (Component~II in  
Fig.~\ref{geometry}). 
It is uncertain whether this polar scattering component is ubiquitous  
in all symbiotic 
stars. \citet{har} showed that a significant fraction 
of symbiotic stars may possess this component, in the  
picture presented by \citet{lee5}. Because the polar component 
is assumed to be 
receding from the emission region, it will mainly affect the red part 
of the H$\alpha$ wings. 
 
With this geometry, we prepare a photon source at the centre, from 
which the ultraviolet photons around 
Ly$\beta$ are injected into the scattering  
region. 
For a given incident photon the corresponding cross-sections of the 
Rayleigh and the Raman scatterings are calculated from the  
Kramers-Heisenberg formula \citep[e.g.,][]{bet,sak}. 
The ratio of the two cross-sections is dependent only upon the  
incident wavelength \citep{lee1}. According to this ratio, we 
decide whether a given scattering event is Rayleigh or Raman in  
the Monte Carlo procedure.

Although the ultraviolet photons may be Rayleigh-scattered several times, 
the scattering regions are assumed to be transparent to 
the optical photons converted by Raman scattering. 
Therefore an incident ultraviolet photon 
suffers a number of Rayleigh scatterings until it is 
Raman scattered or escapes the scattering region as an ultraviolet photon. 
 
We typically inject $10^5$ ultraviolet photons into a bin with  
$\Delta \lambda=0.005\rm\ \AA$ around Ly$\beta$, 
and calculate the polarization state at each scattering. 
Finally we collect Raman scattered photons according to the emergent 
direction cosine of the wave vector with a bin size of $\Delta \mu=0.1$. 
 
\subsection{Computation of Polarization} 
 
Because our scattering geometry is azimuthally symmetric, under the assumption 
that the emission source is not circularly polarized, no circular polarization 
will develop. Hence we only consider linear polarization that may develop 
either in parallel or perpendicular to the symmetry axis. We will distinguish 
the polarization direction by the sign of the polarization. That is, 
we will denote the polarization perpendicular to the symmetry axis by positive 
polarization and parallel polarization by negative polarization. 
 
In the case of single electron atoms including hydrogen, the fine  
structure level splitting is small because of accidental degeneracy. 
Therefore 
any off-resonance scattering is characterized by the same scattering 
phase function as that of the Thomson scattering, which was shown by 
\citet{ste}. Since the H$\alpha$ wings spread over $\ge 1000{\rm\ 
km\ s^{-1}}$, resonance scattering contributes very little  
to the wing formation. 
Hence, in this work, the relevant scattering phase function is that for 
Thomson scattering. 
 
In the computation of linear polarization, we adopt the density 
matrix formalism, in which the polarization state is described by a $2\times 
2$ Hermitian matrix $\rho$. In particular, 
with respect to the symmetry axis which is taken as the  
{\boldmath$z$}-axis in the usual spherical coordinate, 
we choose the polarization basis vectors  
by ${\textit{\boldmath$\epsilon$}}_1=(\sin\phi,-\cos\phi, 0)$, 
${\textit{\boldmath$\epsilon$}}_2=(\cos\theta\cos\phi,\cos\theta\sin\phi,  
\sin\theta)$ for 
a given line photon with wavevector  
${\hat \textit {\boldmath$k$}}=  
(\sin\theta\cos\phi,\sin\theta\sin\phi,\cos\theta)$,  
the linear polarization $p$ is obtained 
from the difference of the main diagonal elements of the density matrix  
$\rho$, 
\begin{equation} 
p=\rho_{11}-\rho_{22}. 
\end{equation} 
 
The relation between the density matrix $\rho'$ of the scattered radiation 
with wavevector  
${\hat \textit {\boldmath$k$}'}= 
(\sin\theta'\cos\phi',\sin\theta'\sin\phi',\cos\theta')$  
and $\rho$ of the incident radiation is given by  
\begin{eqnarray} 
\lefteqn{\rho_{11}'=\rho_{11}\cos^2\Delta\phi 
-\rho_{12}\cos\theta\sin2\Delta\phi 
+\rho_{22}\cos^2\theta\sin^2\Delta\phi,}\\ 
\lefteqn{\rho_{12}'={1\over2}\rho_{11}\cos\theta'\sin2\Delta\phi 
+ \rho_{12}(\cos\theta\cos\theta'\cos2\Delta\phi 
+\sin\theta\sin\theta' \cos\Delta\phi)  
-\rho_{22}\cos\theta(\sin\theta\sin\theta'\sin\Delta\phi  
+{1\over2}\cos\theta\cos\theta'\sin2\Delta\phi),} \nonumber\\ 
\lefteqn{\rho_{22}'=\rho_{11}\cos^2\theta'\sin^2\Delta\phi  
+\rho_{12}\cos\theta'(2\sin\theta\sin\theta'\sin\Delta\phi 
+ \cos\theta\cos\theta'\sin2\Delta\phi)  
+\rho_{22}(\cos\theta\cos\theta'\cos\Delta\phi+\sin\theta\sin\theta')^2,} 
\nonumber 
\end{eqnarray} 
where $\Delta\phi=\phi'-\phi$. 
 
It can be also shown that the angular distribution of the scattered  
radiation for an incident 
photon with ${\hat \textit {\boldmath$k$}}$ and $\rho$ is given by the 
trace of $\rho'(\theta',\phi',\theta,\phi)$ \citep[e.g.][]{lee6}. 
Therefore, once the wavevector ${\hat \textit {\boldmath$k$}'}$ 
of a scattered photon 
is chosen from $\rho'$, the polarization 
is also determined at the same time. It is also noted that the density  
matrix is normalized each time so that it has a unit trace.

\section{Results} 
First, we present two idealized cases in order to obtain a 
qualitative understanding of the polarization development through  
Raman scattering. Next, adopting a simple geometry we calculate the profile 
and the polarization for various geometrical parameters. Then we present 
two applications for more realistic geometries. 
 
\subsection{Infinite Slab} 
\begin{figure} 
\centerline{\includegraphics[width=8cm]{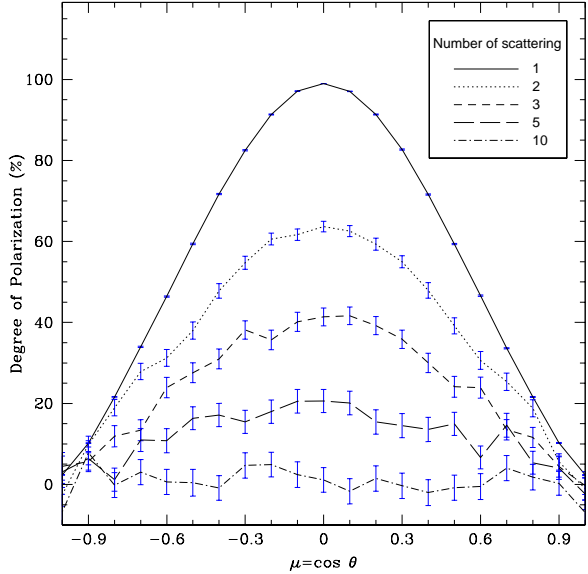}} 
\caption{Polarization of the Raman scattered photons 
for various lines of sight. Ultraviolet photons are normally incident on an 
infinite slab with a finite thickness. 
Various types of lines represent the number of scatterings that  
outgoing photons suffer.  
The polarization is set to be positive when the polarization  
direction is perpendicular to the slab normal. The horizontal 
axis represents the cosine of the angle ($\mu=cos \theta$) between the 
slab normal and the line of sight. The error bars represent  
one standard deviation.} 
\label{num} 
\end{figure} 
\begin{figure*} 
\centerline{\includegraphics{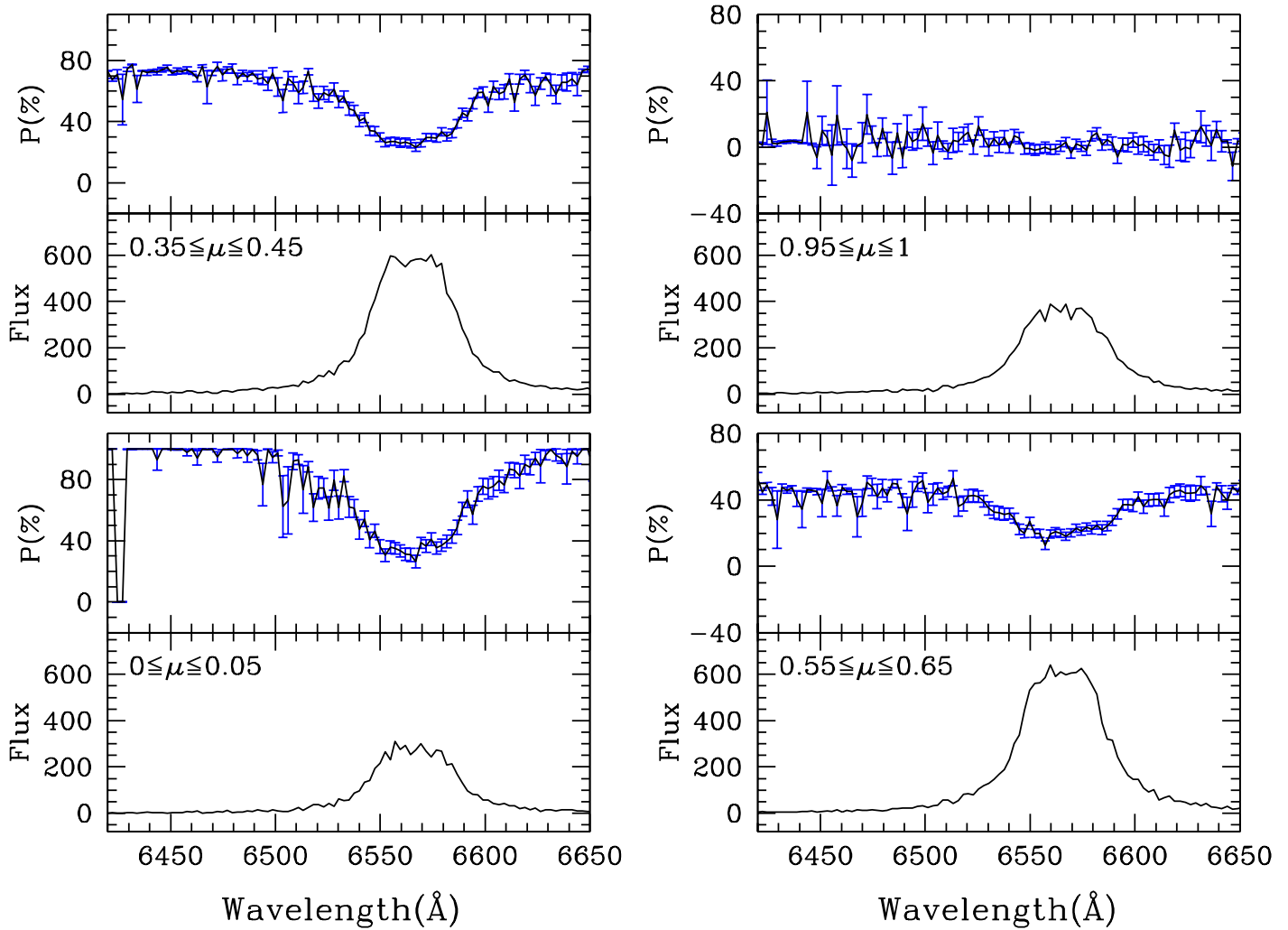}} 
\caption{Profiles and polarization of Raman scattered  
photons for various lines of sight. $10^4$ ultraviolet photons 
are injected normally on to the infinite slab. 
The H~{\tiny{I}} column density is set to be  
$5\times10^{20}\rm cm^{-2}$. The error bars represent one standard deviation.} 
\label{slab} 
\end{figure*} 
To investigate the dependence of the polarization on the number  
of Rayleigh scattering that a photon suffers, we just focus on the  
scattering events, ignoring other effects. An incident photon can escape 
and be counted only if it attains the prescribed number of scattering. 
 
We typically prepare $10^4$ photons which are normally incident on an 
slab with a finite thickness. Here, we assume that the horizontal 
physical dimensions of the slab are infinite. 
We collect only those 
photons that have been scattered the prescribed number of times. 
We present the result in Fig.~\ref{num}, where the horizontal axis 
represents the cosine of angle ($\mu=cos \theta$) between the slab 
normal and the line of sight. The polarization  
steeply decreases as the scattering number increases. 
If a photon is scattered three times, its maximal polarization reduces 
to about 40 per cent, and almost zero polarization is observed 
when the number of scatterings exceeds 10. The general behaviour of 
polarization weakening as a result of increased number of scatterings
can be found in the literature \citep[e.g.][]{ang}. 
 
Next, we present an idealized case of incidence normal to the  
slab in Fig.~\ref{slab}. We also inject $10^4$ ultraviolet 
photons in a bin with  
$\Delta\lambda=0.05\rm\ \AA$ with full width $6\rm\ \AA$ centred 
at Ly$\beta$. They are normally incident to the infinite slab  
lying in the {\boldmath$x-y$} plane. The H~{\tiny{I}} column density is 
set to be $5\times10^{20}\rm cm^{-2}$. 
 
Despite the large column density, 
for an incident ultraviolet photon far from the 
resonance wavelength of Ly$\beta$, 
the scattering region is optically thin and 
hence the Raman conversion rate is very low.  
However, radiation in this wing region is highly polarized up to 100 per
cent
in the equatorial direction, because the scattering number rarely exceeds 2.

As the wavelength goes close to the Ly$\beta$ line centre, the  
cross-section increases steeply. In this case, most incident ultraviolet
photons may suffer a number of Rayleigh scatterings before escape 
by Raman scattering. They escape from the slab 
by the final Raman scattering and contribute to 
the broad H$\alpha$ wing formation. 
The flux near the Ly$\beta$ line centre is typically polarized by  
about 40 per cent in the equatorial direction, which is weaker than in 
the far-wing regions.  The weakening of polarization in this region 
is attributed to the increased Rayleigh scattering numbers before 
escape.

\subsection{Polarization for Various Heights of  
the Cylindrical Scattering Regions} 
\begin{figure*} 
\centerline{\includegraphics{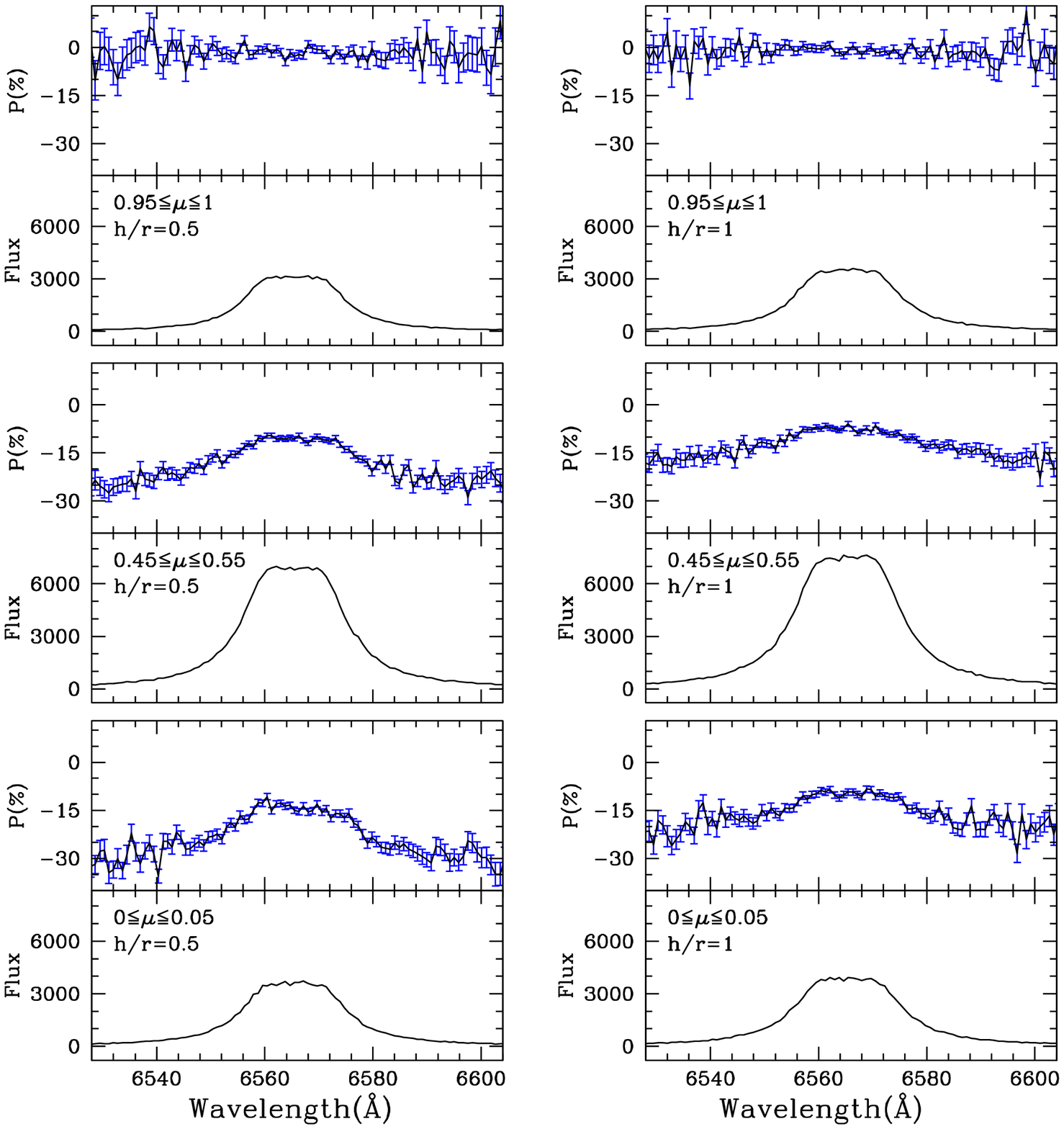}} 
\caption{Profiles and polarization of broad H$\alpha$ 
wings for various lines of sight. Here we consider only Component~I in 
Fig.~\ref{geometry} with $h\equiv h_1=h_2=h_3$. 
The height-to-radius ratios $h/r$ of the H~{\tiny{I}} cylindrical shell  
are 0.5 and 1 for left- and right-hand panels, respectively. 
$10^5$ ultraviolet photons in a bin with $\Delta\lambda=0.005\rm\ \AA$ 
are injected to the H~{\tiny{I}} component~I. 
The homogeneous H~{\tiny{I}} column density is set to be  
$10^{20}\rm cm^{-2}$. The error bars represent one standard deviation.} 
\label{result1} 
\end{figure*} 
\begin{figure*} 
\centerline{\includegraphics{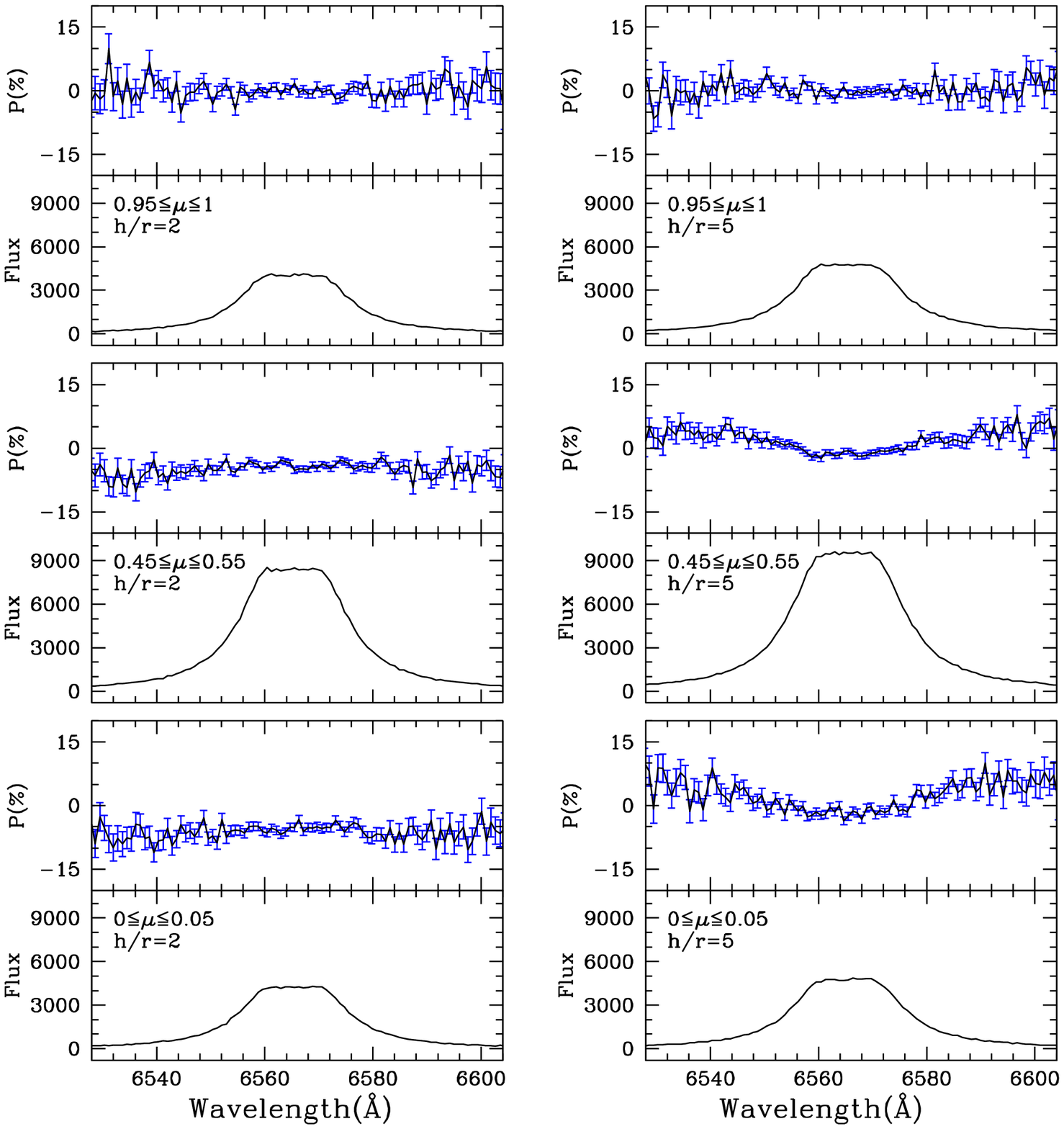}} 
\caption{Profiles and polarization of broad H$\alpha$ 
wings for various lines of sight. The 
height-to-radius ratios of the H~{\tiny{I}} 
cylinder are 2 and 5 for left- and right-hand
panels, respectively. Other parameters are the same as described in 
Fig.~\ref{result1}} 
\label{result2} 
\end{figure*} 
We assume that the H~{\tiny{I}} number density is constant throughout 
the scattering region of Component~I in Fig.~\ref{geometry} 
($h\equiv h_1=h_2=h_3$), and obtain the profile and 
polarization of broad H$\alpha$ wings, changing the ratios $h/r$ of  
height to radius.  
In this work, just for simplicity and with a lack of detailed information on
the circumstellar matter distribution, we assume that the thickness of  
the H~{\tiny{I}} cylindrical shell is much smaller than the radius.  
The profiles and polarization of the broad H$\alpha$ wings 
for various lines of sight are shown for several ratios 
in Figs.~\ref{result1} and ~\ref{result2}. 
 
The radiation emergent in the polar direction is almost unpolarized 
because of the symmetry, which provides a check of our code. 
When the observer is in the equatorial direction, 
the Raman scattered radiation exhibits negative polarization.  
As the height-to-radius ratio approaches zero, the scattering region 
is regarded as a circular ring, which yields the strongest polarization,
33 per cent,
in the direction parallel to the symmetry axis of the circular ring.  
Since photons falling on to far-wing regions are scattered fewer times 
than those contributing to near-wing regions, the polarization in far-wing 
regions 
is much stronger than that near the H$\alpha$ centre. 
Therefore it is particularly notable that the Raman scattering wings are  
characterized by weak polarization 
near the centre and strong polarization at far-wing regions, in the  
direction of the symmetry axis. 
 
As the height increases relative to the radius, for a given line of 
sight, there can be a broad range of scattering angles. Therefore 
a superposition of electric vectors associated with the scattered 
radiation lessens the overall polarization. In Fig.~\ref{result2},  
it is apparent 
that the polarization weakens as $h/r$ increases. When $h/r=2$, 
we obtain H$\alpha$ wings polarized up to 10 per cent observed 
from the equatorial direction, which is much smaller than 30 per cent 
obtained in the case of $h/r=0.5$. 
 
This trend will continue until the geometry is approximately regarded 
as an infinite cylinder. In this case, when the cylinder 
has enough thickness, emergent Raman-scattered wings will be 
contributed to equally from all directions of incidence, and will be 
completely unpolarized.  We increase the ratio $h/r$ to as high as 5,  
which corresponds to the case that ultraviolet radiation from 
the emission source  
with $\theta>10^\circ$ will be incident on to the scattering region. 
As is shown in upper panels of Figs.~\ref{result1} and \ref{result2},  
we obtain negligible polarization in this case for all lines of sight. 
 
However, we also see the weakly polarized parts in far blue and red regions 
for the case $h/r=5$. 
This polarization, developed in the direction perpendicular to the  
symmetry axis, is attributed to the limb effect. That is, the 
cylindrical shell is optically thin in the direction normal to the shell 
for far blue- or red-wing photons. Therefore, only 
a small fraction $\tau_w$ of photons incident in this direction will 
be scattered, where $\tau_w$ is the Raman scattering optical depth in 
this direction. However, when the incident photon has the wavevector 
${\hat \textit {\boldmath$k$}}$ with  
${\hat \textit {\boldmath$k$}}\cdot {\hat \textit {\boldmath$z$}}= 
\mu\simeq 1$, the fraction of scattered photons increases to  
${\tau_w}/{\sqrt{1-\mu^2}}$ which is much larger than $\tau_w$. 
This limb effect strengthens the positive polarization component and 
results in polarization in the perpendicular to the symmetry axis. 
 
In the cylinder scattering model, we may summarize by saying that
the polarization develops in the direction parallel to the symmetry 
axis, and that the polarization is weaker near the H$\alpha$ centre,  
because of large number of the Rayleigh scatterings, 
than it is in the far-wing 
regions. 
 
\subsection{Hybrid Model of Cylindrical Shells} 
\begin{figure*} 
\centerline{\includegraphics{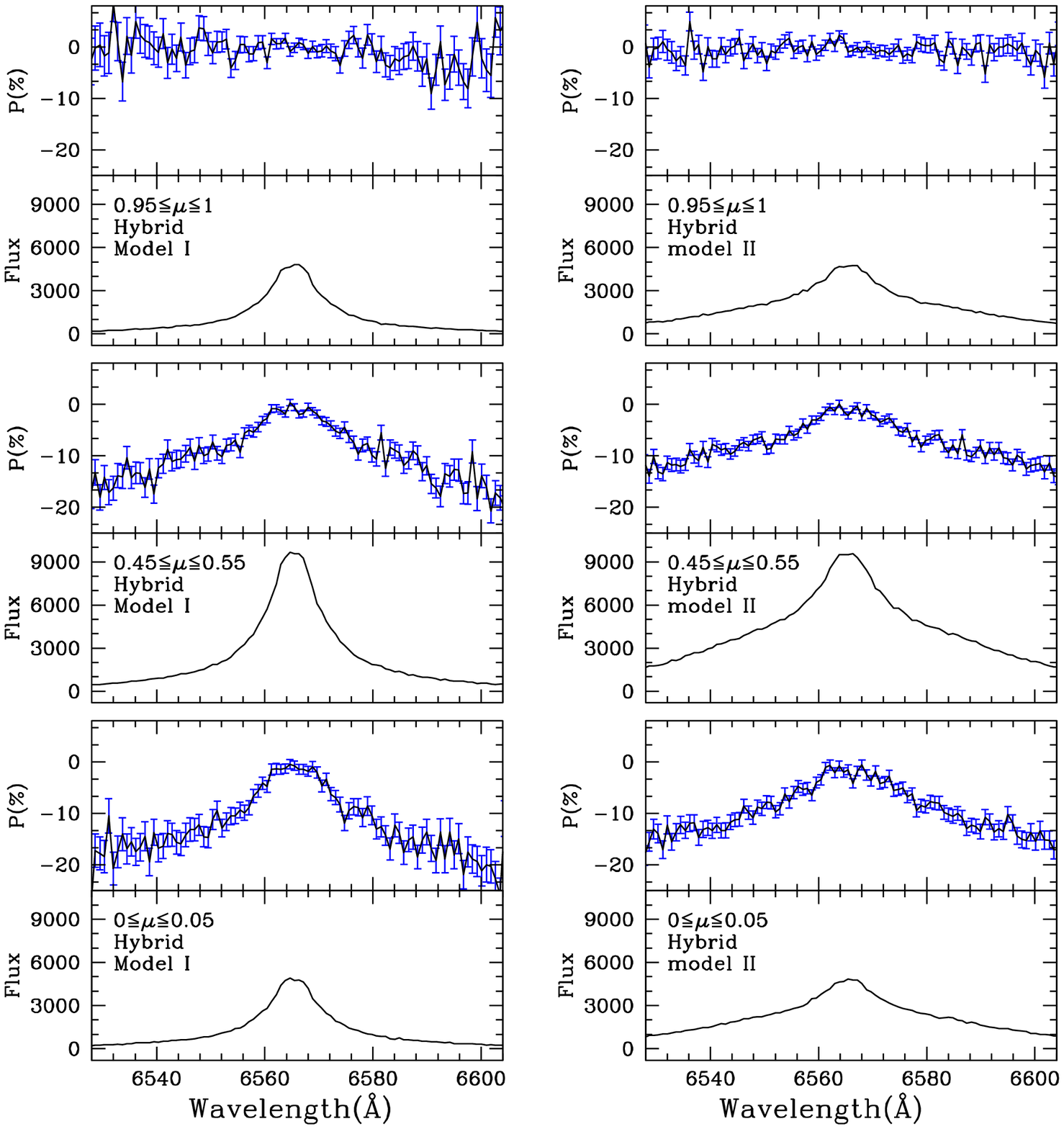}} 
\caption{Profiles and polarization of broad H$\alpha$ wings for  
the hybrid model. We consider only Component~I in Fig.~\ref{geometry}. 
The H~{\tiny{I}} column densities are set to be  
$10^{21}, 10^{20}$ and $10^{19} \rm \ cm^{-2}$ from the equatorial region 
with $h_1, h_2, h_3$=0.1, 0.5 and 5 in units of the radius $r$ 
for the Hybrid Model~I, and 0.5, 1 and 
5 for the Hybrid Model~II. Other parameters are the same as described 
in Fig.~\ref{result1}.} 
\label{hybrid} 
\end{figure*} 
 
We present a hybrid model of Component~I in Fig~\ref{geometry}  
consisting of three stratified layers ($h_1<h_2<h_3$) 
with different H~{\tiny{I}} column densities. 
In this model, the density is highest in the equatorial 
region and decreases as the latitude increases. This may mimic the 
matter distribution in the circumstellar region more realistically.   
This hybrid model can be viewed as a linear superposition which is the 
solid-angle-weighted sum of the previous models with various ratios and  
homogeneous hydrogen number density.  
 
We consider two different cases denoted by Model~I and Model~II. In 
Model~I, we set $h_1 = 0.1 r$, $h_2=0.5 r$ and $h_3= 5r$;
in Model~II, we set $h_1=0.5r,\ h_2=r$ and $h_3=5r$. In both cases, the 
H~{\tiny I} column densities of the stratified layers are chosen  
to be $N_{HI}=10^{21},\ 10^{20}$ and $10^{19}{\rm\ cm^{-2}}$, 
respectively.

We obtain similar patterns in polarization, 
which is weak near the H$\alpha$ centre and becomes strong in 
far-wing regions. These polarization patterns are also attributed to 
larger scattering numbers for photons near the line centre than 
those for far-wing photons. 
 
It is particularly notable that a flat profile near the line centre 
no longer appears. This is because of the contribution made by the small  
column density regions at high latitudes. These regions are 
optically thin for most of incident ultraviolet photons, and hence yields 
profiles proportional to the scattering cross-section which is very 
steep near the line centre. In Model II shown in the right-hand panel of 
Fig.~\ref{hybrid}, the profile is fairly extensive because of the 
much larger column 
density than in the case of Model I shown in the left-hand panel of the 
same figure. 
 
\subsection{Raman Scattering from the Receding Polar Components} 
\begin{figure*} 
\centerline{\includegraphics{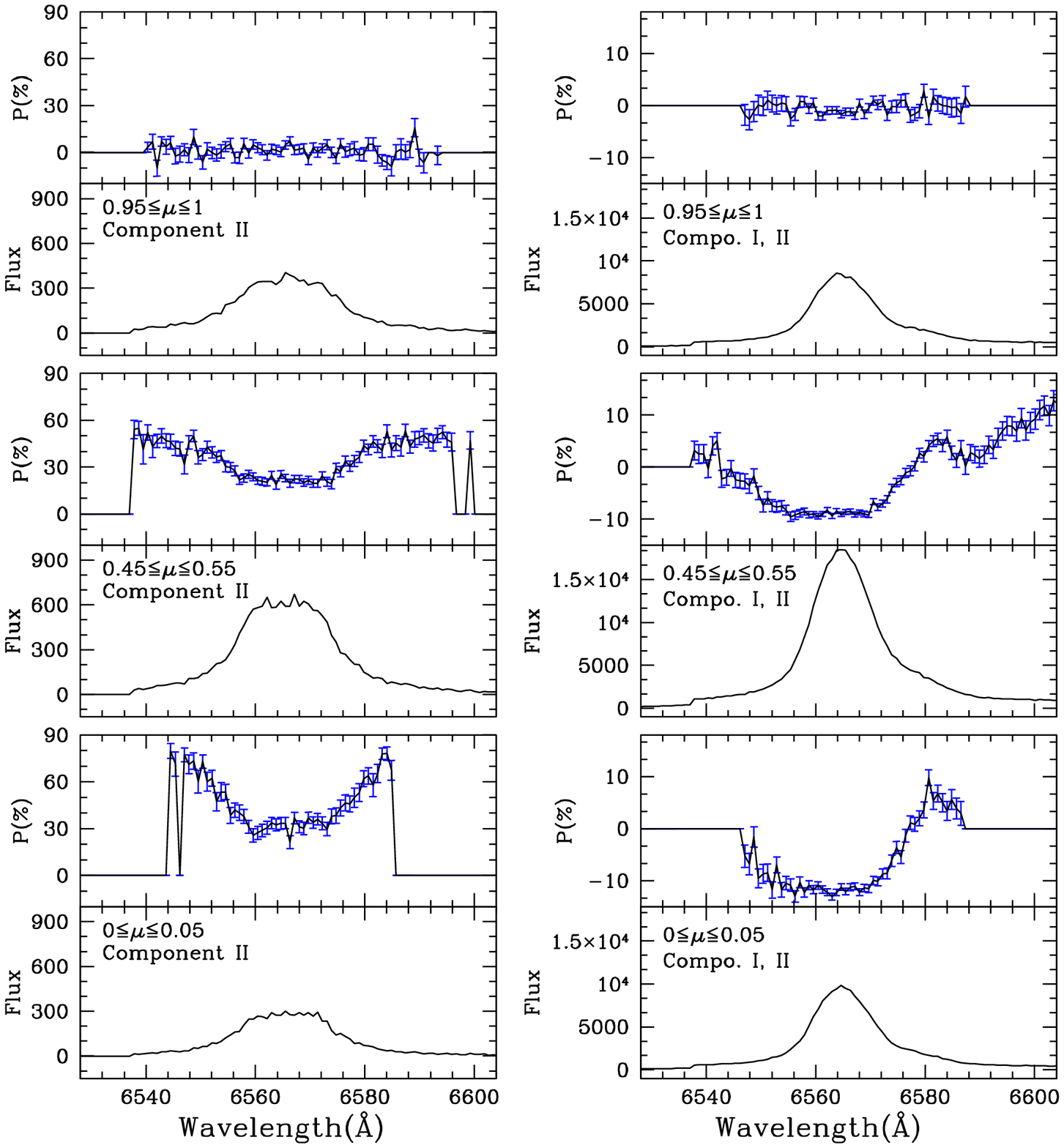}} 
\caption{Profiles and polarization of broad H$\alpha$ 
wings for both components. We set the recession velocity of the 
scattering region of Component~II to be $v_p=100\rm\ km\ s^{-1}$,  
the half-opening angle to be 20\degr\, and the ratio of Component~I 
($h\equiv h_1=h_2=h_3$) to be $h/r=0.5$.  
$10^5$ ultraviolet photons in a bin with $\Delta\lambda=0.005\rm\ \AA$ are  
injected into only Component~II with H~{\tiny{I}} column density 
$10^{20}\rm\ cm^{-2}$ in the left-hand panel, and into both components  
(H~{\tiny{I}} column density $10^{20}$, $10^{23}\rm\ cm^{-2}$ for  
Component~I and Component~II, respectively) in the right-hand
panel. We omit to denote the polarization when the photon flux is  
negligibly small. The error bars represent one standard deviation.} 
\label{com} 
\end{figure*} 
In this subsection, we consider the Raman scattering from the receding 
polar scattering region of  
Component~II, the existence of which is implied by \citet{har}. 
In this work, we simply assume that this component is moving away 
from the emission source along 
the symmetry axis of the cylindrical shells considered in the previous 
subsections. 
 
In the left-hand panel of Fig.~\ref{com} 
we show the polarization of H$\alpha$ wing 
photons Raman-scattered in the receding polar 
region. We set the recession velocity of the polar component 
to be $v_p=100{\rm\ km\ s^{-1}}$. Since the symmetry axis and the scattering 
plane make a very small angle, the polarization develops in the direction 
perpendicular to the symmetry axis, or in our formalism we obtain positive 
polarization. For an observer in the equatorial direction, the scattering 
angle is nearly 90$^\circ$, and therefore almost complete polarization 
is obtained. 
 
The wing flux is proportional to the scattering cross-section 
and the incident UV flux in the optically thin limit. In the observer's rest 
frame, the scattering cross-section peaks at the recessional velocity of 
the polar component $v_p$, and therefore we obtain a small peak at this 
velocity. Hence the resultant profile is obtained effectively by shifting 
the wing flux that is formed from Raman scattering in a static medium. 
Owing to the incoherency of Raman scattering, the wing flux peaks 
at 6578 \AA.  This is interesting in that the symbiotic stars 
RR~Tel and He~2-106 appear to exhibit a small and broad bump around 
6580 \AA\ just blueward of [N~II]~6584 feature \citep{lee4}. 
 
In the right-hand panel of Fig.~\ref{com},  
we present the Raman scattered wing fluxes obtained from 
combinations of a static cylindrical shell 
and a receding polar component. 
Since the two scattering components provide opposite senses of polarization, 
the resultant polarization is obtained by the algebraic sum 
of the two polarized flux components divided by the total flux.  
Therefore, in the blue wing, where the polar 
component contributes little, we obtain polarization in the 
direction parallel to the symmetry axis. However, in the red wing, the polar 
component predominantly contributes to the final polarization, yielding 
perpendicular polarization. 
 
\section{Discussion and Summary} 
We have investigated the polarization of H$\alpha$ wings formed through 
Raman scattering of Ly$\beta$. Two kinds of scattering regions are 
considered. The first one is a static cylindrical shell. From 
this component, the polarization of H$\alpha$ wings develops  
in the direction parallel 
to the cylinder axis. It is particularly notable that stronger 
polarization is obtained in far-wing regions than near the line 
centre.  Near the line centre, the scattering cross-section 
is larger, and therefore Rayleigh scattering occurs 
more frequently than in the far-wing regions. The weaker 
polarization in the near-wing regions is attributed to the larger 
Rayleigh scattering numbers than in the far-wing regions. 
 
We consider the second scattering component which is located in the 
polar regions and moving away from the ultraviolet emission source 
with a recessional velocity $v_p=100{\rm\ km\ s^{-1}}$. The physical 
nature of this component is not certain, but the existence of this 
component is apparent from the spectropolarimetric observations 
presented by \citet{har}. This component produces 
polarization in the direction perpendicular to the symmetry axis. 
Since the scattering angle is almost $90^\circ$, nearly complete 
polarization is obtained. Since the scattering component is moving 
away from the source, the wing flux peak appears at $v=6.4 v_p$, 
or around 6578 \AA\ with our choice of $v_p=100{\rm\ km\ s^{-1}}$. 
Therefore, this polar component will produce a small bump in the wing profile 
which is strongly polarized in the direction perpendicular to the 
symmetry axis. 
 
When the both scattering components co-exist, we may observe 
the polarization flip, where the blue part is characterized by 
parallel polarization and the red part is polarized in the 
perpendicular direction. Therefore, if the scattering region 
is composed of a cylindrical shell 
subtending a fairly large solid angle with sufficient column density 
and a small receding polar component, the overall wing profile 
is determined by the cylinder component and the polarization 
will be predominantly contributed by the polar component in the red part. 
Therefore in this case, we mainly obtain a profile that is nearly 
proportional to $\Delta\lambda^{-2}$ and strongly polarized in the 
red part in the direction perpendicular to the symmetry axis. 
 
This is very interesting, considering the spectropolarimetric 
observation of the symbiotic star BI~Crucis by \citet{har3}, 
who reported the polarized H$\alpha$ wings in the direction 
perpendicular to the symmetry axis. They also noted that the polarization 
is very strong in the red part. In their spectropolarimetric data, no 
polarization flip is apparent. They interpreted their data using the 
assumption that the wings are formed through electron scattering. 
However, their data appear to be qualitatively consistent with 
the hypothesis that the H$\alpha$ wings are formed by 
Raman scattering in a fairly extensive cylindrical shell and 
a receding polar component. 
 
\citet{lee7} proposed that electron scattering of monochromatic 
line radiation may leave polarization structures. The radiative transfer 
process through Thomson scattering is diffusive both in real space 
and frequency space. This implies that photons that suffer more electron  
scattering tend to fall in both far- and near-wing regions, whereas 
those with fewer numbers of scatterings contribute to the formation of 
near-wing regions. Since polarization becomes weaker as the scattering number 
increases, we obtain weaker polarization in the far-wing regions in 
Thomson scattering media. Therefore spectropolarimetry can be an important 
tool to investigate the wing formation mechanism. 
 
The Ly$\beta$ photon source was considered to be point-like. In 
reality, this photon source will be extended and even partly coincide 
with the scattering region. This will most likely lower the resulting 
polarization. The H$\alpha$ line flux will be strongly diluted 
by photons from the emission nebula, originating predominantly from 
the recombination process. This contribution can be dominant in the line 
centre and it will lower the resulting polarization considerably. 
Other factors that may affect the polarization of Raman-scattered H$\alpha$ 
wings and have not been considered in this work may include dust and 
the kinematics 
of the cylindrical shell component. Currently, only a few 
spectropolarimetric observations of H$\alpha$ wings can be found in the 
literature, and little is known about the formation of H$\alpha$ profiles 
seen in symbiotic stars. We hope that more observations will provide  
constraints on various theoretical models of H$\alpha$ and other Balmer 
series line formation.    
 
\section*{Acknowledgments} 
We are very grateful to the referee, H. Schmid, whose comments greatly 
improved the presentation of this paper. 
This work was the result of research activities `The Astrophysical Research Center for the Structure and Evolution of the Cosmos' 
supported by Korea Science \& Engineering Foundation.

\bsp \\ 
 
\end{document}